# Rapid generation of thermal-safe test schedules

Paul Rosinger, Bashir Al-Hashimi*
University of Southampton
School of Electronics and Computer Science
Southampton, SO17 1BJ, UK
{pmr,bmah}@ecs.soton.ac.uk

Krishnendu Chakrabarty
Dept. of Electrical and Computer Engineering
Duke University
Durham, NC 27708
krish@ee.duke.edu

**Abstract**

*Overheating has been acknowledged as a major issue in testing complex SOCs. Several power constrained system-level DFT solutions (power constrained test scheduling) have recently been proposed to tackle this problem. However, as it will be shown in this paper, imposing a chip-level maximum power constraint doesn't necessarily avoid local overheating due to the non-uniform distribution of power across the chip. This paper proposes a new approach for dealing with overheating during test, by embedding thermal awareness into test scheduling. The proposed approach facilitates rapid generation of thermal-safer test schedules without requiring time-consuming thermal simulations. This is achieved by employing a low-complexity test session thermal model used to guide the test schedule generation algorithm. This approach reduces the chances of a design re-spin due to potential overheating during test.*

## 1. Introduction

Considering power consumption during test is important because recent industrial experience has shown that scan testing in some designs may consume almost 30X of peak power over its normal operation mode [11]. A difference of such a magnitude can easily lead to permanent damage to the device under test due to overheating, or, lead to reliability failures due to electro-migration.

Recent research has addressed the problems associated with the excessive power dissipation during test both at core and at system level. Core level solutions include improved ATPG algorithms [13], pattern ordering methods [3], and scan chain and clocking scheme modifications [10]. Existing system level solutions consist of various power-constrained test scheduling algorithms [2, 6, 7, 5, 4, 1, 9, 8] which limit the concurrency of the core tests based on a chip-level maximum allowable power limit. This paper focuses on system level tackling of overheating during test.

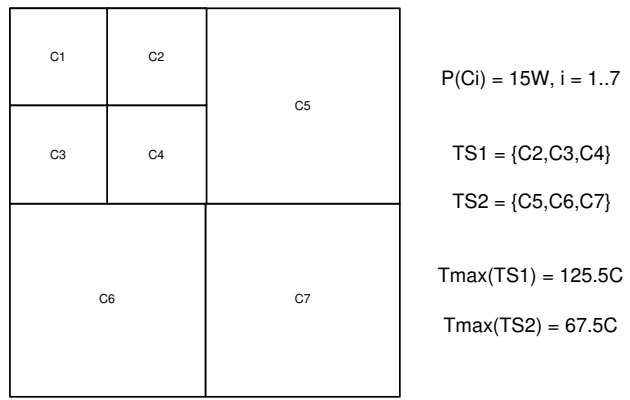

**Figure 1. Effect of power density variations on**

Silicon die hot spots resulted from localised heating occur much faster than chip-wide overheating due to the non-uniform spatial on-die power distribution. Accepting the assumption of non-uniform spatial power distribution, we belive that constraining the maximum

*The first two authors would like to acknowledge the Engineering and Physical Sciences Research Council (EPSRC) for funding this work under grant no. GR/S05557. The authors also wish to thank the anonymous reviewers for their suggestions which helped improve the quality of the paper.

chip-level power dissipation is not effective in avoiding localised overheating. To demonstrate this, we provide the following motivational example based on a hypothetical system shown in Figure 1. This system is a typical example of non-uniform power distribution: cores with different sizes dissipate the same amount of power. In a power constrained test scheduling scenario, for a power constraint of 45W the two test sessions TS1 = {C2,C3,C4} and TS2 = {C5,C6,C7} would be both accepted by the test scheduling algorithm. However, thermal simulation results show a large discrepancy in terms of maximum temperature between the two test sessions, 125.5°C for TS1 vs. 67.5°C for TS2. This difference is mainly because the power density (power per unit of area) varies significantly from one core to another. For example, the power density of core C2 is 4 times higher than that of C5.

This means that in order to efficiently avoid hot spots without unnecessarily reducing test concurrency, the spatial and temporal non-uniform thermal behaviour of the chip under test must be addressed directly, i.e. by validating the generated test schedules through thermal simulations. This paper proposes a new approach for dealing with overheating during test, by embedding thermal awareness into test scheduling. To the best of our knowledge, this is the first investigation where test session thermal models are employed to guide the test schedule, rather than chip-level power constraints [2, 6, 7, 5, 4, 1, 9, 8].

## 2. Proposed Test Session Thermal Model

Accurate thermal simulation can be very time consuming for complex chips, therefore it is essential to keep the number of test schedule re-generation due to thermal violations to a minimum. In order to achieve this, we propose a low-complexity test session thermal model used to guide the test schedule generation. This reduces the amount of accurate thermal simulations necessary for obtaining a thermal-safe test schedule.

The heat generated during a test session by an active core can be transfered away from the core through its lateral neighbourhood and through the heat spreader placed above the silicon die. Limited lateral heat spreading elevates the core temperature as the only available heat release path remains the vertical one, i.e. through the heat spreader. The basic idea behind the proposed thermal-aware test schedule generation approach is to maximise the amount of heat which can be dissipated through the lateral neighbourhood of each active core in a test session. The proposed test session thermal model captures at core granularity level the main heat transfer paths originating at the cores tested in a given test session. This model is inspired from the RC-equivalent architecture-level thermal model proposed in [12] which exploits the duality between the thermal and electric domains by modelling an IC as a network of thermal resistances and thermal capacitances. In the RC thermal model, each core is represented as a node in the RC network and thermal adjacency is modelled with thermal RC pairs connecting the corresponding nodes.

To adapt the generic RC-equivalent model proposed in [12] to the specific needs of the thermal-aware test schedule generation, we have made the following modifications:

1. Only steady-state temperatures are considered as they represent upper bounds for the transient thermal profiles of individual cores. Therefore, only the thermal resistances of the generic RC model are used.

2. The heat transfer between two cores tested concurrently is considered to be negligible, and hence the thermal resistances between cores tested in the same test session are removed. This is a valid assumption because, the amount of heat exchanged by two adjacent objects depends on their temperature difference, and the temperature difference between two active cores is less than that between an active and passive core.

3. The cores which are passive in the test session under consideration, are assumed to be thermally grounded, i.e. their temperature is equal to the ambient temperature and fixed for the entire duration of the test session.

These modifications of the thermal model simplify the thermal-aware test schedule generation while still producing effective solutions, as demonstrated by the experimental results reported in Section 4.

2Proceedings of the Design, Automation and Test in Europe Conference and Exhibition (DATE'05)
1530-1591/05 $ 20.00 IEEE

The following example illustrates the proposed test session thermal model. For the layout configuration shown in Figure 2, we assume the test session under consideration consists of the tests for cores 2,4 and 5. The white arrows show the lateral paths available for moving the heat away from the active cores. Figure 3 shows the thermal resistive model (modification 1) corresponding to this test session. The thermal resistances between the nodes corresponding to active cores are omitted (modification 2), while all remaining thermal resistances connect the active core nodes to the thermal ground (modification 3). The equivalent test session thermal model shown in Figure 4 will be used to guide the thermal-aware test schedule generation as follows. A small equivalent thermal resistance associated with an active core means good heat exchange between the core and the ambient, which predicts a low core temperature during test. On the other hand, a large equivalent thermal resistance associated with an active core means poor heat exchange with the ambient, and therefore signals a potential hot spot during test.

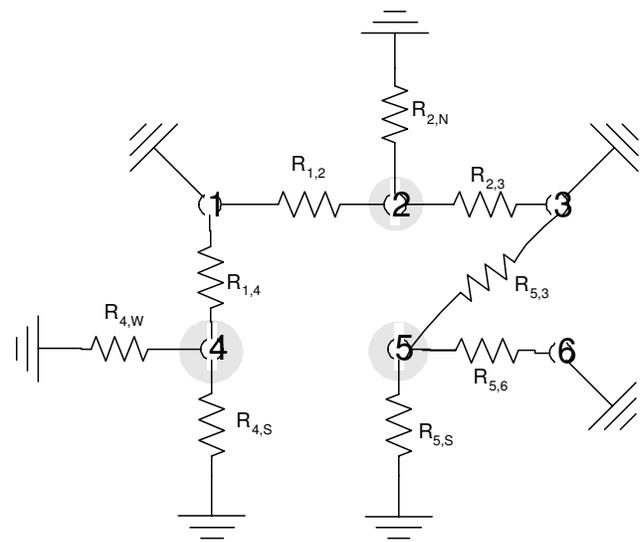

**Figure 3. Test session thermal model of example in Figure 2**

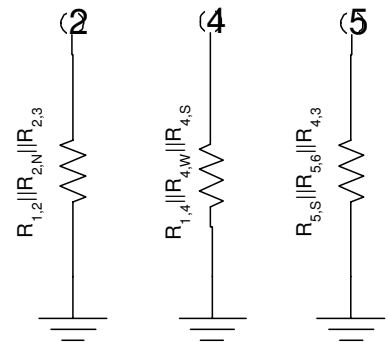

**Figure 4. Equivalent test session thermal model**

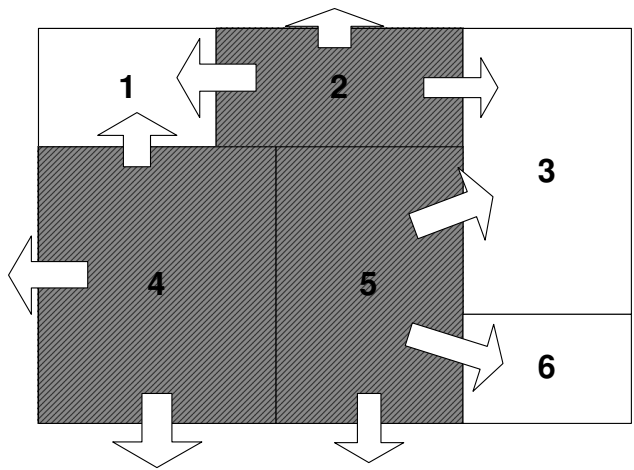

**Figure 2. Test session example**

Power dissipation differs from core to core, therefore, we are introducing the core thermal characteristic(TC) with respect to a given test session TS defined as: $TC_{TS}(i) = P(i) \times R_{th}(i)$, where P(i) is the average power dissipation for core $i$ and $R_{th}(i)$ is the equivalent thermal resistance associated with core $i$ with respect to TS. TC provides a normalised means for selecting the appropriate core to be added to a test session. This means that a core with a poor heat exchange configuration (large $R_{th}$) but low power dissipation has comparable chances to be assigned to a test session with a core which would exhibit good heat exchange with the ambient but has higher power dissipation.

The proposed thermal-aware test scheduling algorithm, which will be detailed in the next section, is driven by the test session thermal characteristic (STC), defined as follows:

STC(TS) = $\max_{C_i \in TS}(TC_{TS}(i) \times P(i) \times W(i))$,

where W(i) is a weight associated with core i, and initially set to 1.

## 3. Thermal aware test schedule generation

The pseudocode for the proposed thermal-safe test schedule generation flow is given in Algorithm 1. The inputs to the algorithm are the set of cores(S) of the targeted system, the maximum allowable temperature(TL) and the maximum test session thermal characteristic limit (STCL). In the first stage (lines 1-7), the algorithm verifies if the temperature constraint is not violated by individual cores. For this, a purely sequential test schedule, i.e. only one core per test session, is simulated. If a thermal violation occurs, then it is either fixed by redesigning the test infrastructure of the core, or the temperature limit (TL) is increased. Once all individual temperature violations have been eliminated, the algorithm proceeds to the actual test schedule generation. New cores are added to an empty test session TS until STC(TS) exceeds the user-specified STCL(lines 9-15). Once no more cores can be added to TS, TS is simulated and the maximum temperature for each core in TS is compared with TL (line 19). If any thermal violation is detected at this stage, TS is discarded and the weight W of all cores which violated TL is increased. This is done in order to make them less likely to be added to a "busy" test session. If no thermal violation was detected, TS is added to the test schedule. The algorithm continues with a new session (line 9) until all cores have been scheduled.

## 4. Experimental results

In order to validate the efficiency of the proposed thermal-aware test scheduling approach, we have performed a set of experiments based on the Compaq Alpha 21368 floorplan from [12]. The floorplan consists of 15 individual cores describes in terms of their size and position within the floorplan. The test power dissipationvalues used for these cores were ranging from 1.5X to 8X their power dissipation during normal operation. The experiments focus on two categories of results: the length of the generated test session and the simulation effort. By simulation effort we mean the amount of test session time which needs to be simulated until a thermal-safe test schedule is reached. In our experiments we have used the HotSpot tool [12] to

**Algorithm 1** Thermal-safe test schedule generation

**INPUT:** $S$, the core set of the targeted system
TL = max. allowable temperature
STCL = session thermal characteristic limit
**OUTPUT:** Thermal-safe schedule as a list of
thermal-safe test sessions

```
1  FOR EACH C_i ∈ S
2      simulate( C_i )
3      BCMT(i) = MaxTemp( C_i )
4      IF BCMT(i) ≥ TL
5          fix core-level thermal violation OR increase TL
6      END IF
7  END FOR
8  A = { C_i | C_i ∈ S }
9  TS = ∅
10 FOR EACH C_i ∈ A
11     TS1 = TS ∪ C_i
12     IF STC(TS1) ≤ STCL
13         TS = TS1
14     END IF
15 END FOR
16 simulate( TS )
17 ValidSession = TRUE
18 FOR EACH C_i ∈ TS
19     IF MaxTemp( C_i ) ≥ TL
20         W_i = W_i × 1.1
21         ValidSession = FALSE
22     END IF
23 END FOR
24 IF ValidSession
25     add TS to the test schedule
26     A = A - TS
27 END IF
28 IF A ≠ ∅ GO TO LINE 9; END IF
29 DONE
```

perform thermal simulations, however other IC thermal simulation tools could be used just as well.

Figure 5 shows the effect of the session thermal characteristic limit (STCL) on the length of the generated test schedules and on the required simulation effort for TL={145°C, 155°C, 165°C}. As can be seen, relaxed (large) STCL values lead to short test schedule at the expense of a significant simulation effort.

For example, when TL=145°C, a STCL value of 100 leads to a 3 second test schedule, but required a cumulated 26 seconds of test session time to be simulated until a thermal-safe schedule could be identified. The high simulation effort required is due to a large number of thermal violations (line 19 in Algorithm 1) which lead to additional test session generation iterations. As the STCL becomes tighter (smaller value), we notice an increase in the length of the generated test schedules, however the simulation effort involved is much lower compared to the previously discussed case. for very tight constraints (STCL $\leq$ 30), the simulation effort involved equals the length of the generated test schedule, i.e. a thermal-safe test schedule was identified from the first attempt, hence no additional simulations due to thermal violations were required. Figure 5 also shows that as TL in increased, the test schedules get shorter as more test could be assigned to the same test session without violating the thermal constraint. Also the simulation effort decreases because thermal-safe schedules are easier to generate under a more relaxed thermal constraint.

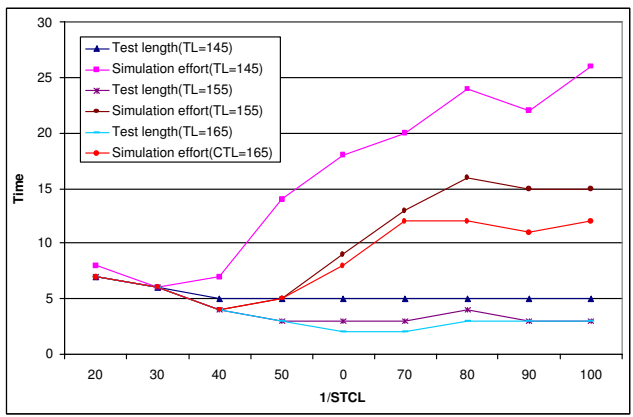

**Figure 5. Test schedule length and simulation effort vs. the session thermal characteristic limit**

Table 1 reports a more extensive set of results for TL in the 145-185°C range. In addition to the test schedule length and simulation effort, the simulated maximum temperature for the entire test session is also reported. As it can be seen, the maximum temperature approaches TL especially for very short test schedules, as they make better use of the temperature allowance.

For example, when TL=150°C, the maximum temperature for a 7 second test schedule was almost 6°C below TL, while for a 4 second test session, the maximum temperature was less than 1°C below TL. These results also show that depending on the STCL, for the same TL, reductions up to 3.5X in test schedule length can be obtained. Another interesting thing to be noted is the fact that for high TL and low STCL, the simulated maximum temperature can be up to 35°C below TL. For example, for TL=185, and STCL=30, the maximum temperature corresponding to the 6 second test stays under 145°C . This shows that in these cases, the STCL constraint is much stronger than TL.

## 5. Conclusions

In this paper we have proposed an approach for test schedule generation guided by a test session thermal model. To the best of our knowledge this is the first investigation targeting the overheating during test based on a thermal model of the test session, rather than imposing a chip-level power constraint. We belive a thermal-aware test scheduling approach is more effective than power constrained test scheduling approached, because of the known low correlation between silicon temperature and power dissipation. As demonstrated by the experimental results, this approach generates thermal-safe schedules, given a maximum temperature limit, while keeping the necessary computational effort to a minimum. Moreover, the proposed approach allows exploration of more efficient solution at the expense of longer thermal simulation times through a user selectable parameter.

| Thermal limit TL °C | STCL | Test schedule length sec | Simulation effort sec | Max. temperature °C |
|---|---|---|---|---|
| 145 | 20 | 7 | 8 | 144.29 |
| 145 | 30 | 6 | 6 | 144.29 |
| 145 | 40 | 5 | 7 | 144.51 |
| 145 | 50 | 5 | 14 | 144.00 |
| 145 | 60 | 5 | 18 | 144.00 |
| 145 | 70 | 5 | 20 | 144.00 |
| 145 | 80 | 5 | 24 | 144.00 |
| 145 | 90 | 5 | 22 | 144.51 |
| 145 | 100 | 5 | 26 | 144.00 |
| 150 | 20 | 7 | 8 | 144.29 |
| 150 | 30 | 6 | 6 | 144.29 |
| 150 | 40 | 4 | 4 | 149.12 |
| 150 | 50 | 4 | 6 | 147.54 |
| 150 | 60 | 4 | 15 | 149.20 |
| 150 | 70 | 4 | 14 | 147.80 |
| 150 | 80 | 4 | 19 | 149.20 |
| 150 | 90 | 4 | 18 | 149.31 |
| 150 | 100 | 4 | 17 | 149.38 |
| 155 | 20 | 7 | 7 | 150.85 |
| 155 | 30 | 6 | 6 | 144.29 |
| 155 | 40 | 4 | 4 | 149.12 |
| 155 | 50 | 3 | 5 | 154.91 |
| 155 | 60 | 3 | 9 | 154.40 |
| 155 | 70 | 3 | 13 | 153.20 |
| 155 | 80 | 4 | 16 | 154.40 |
| 155 | 90 | 3 | 15 | 153.51 |
| 155 | 100 | 3 | 15 | 154.40 |
| 160 | 20 | 7 | 7 | 150.85 |
| 160 | 30 | 6 | 6 | 144.29 |
| 160 | 40 | 4 | 4 | 149.12 |
| 160 | 50 | 3 | 5 | 154.91 |
| 160 | 60 | 4 | 12 | 154.40 |
| 160 | 70 | 3 | 13 | 153.20 |
| 160 | 80 | 3 | 14 | 158.92 |
| 160 | 90 | 3 | 11 | 157.83 |
| 160 | 100 | 3 | 12 | 159.74 |
| 165 | 20 | 7 | 7 | 150.85 |
| 165 | 30 | 6 | 6 | 144.29 |
| 165 | 40 | 4 | 4 | 149.12 |
| 165 | 50 | 3 | 5 | 154.91 |
| 165 | 60 | 2 | 8 | 161.69 |
| 165 | 70 | 2 | 12 | 161.69 |
| 165 | 80 | 3 | 12 | 164.48 |
| 165 | 90 | 3 | 11 | 158.73 |
| 165 | 100 | 3 | 12 | 161.14 |
| 170 | 20 | 7 | 7 | 150.85 |
| 170 | 30 | 6 | 6 | 144.29 |
| 170 | 40 | 4 | 4 | 149.12 |
| 170 | 50 | 3 | 3 | 169.61 |
| 170 | 60 | 2 | 8 | 161.69 |
| 170 | 70 | 3 | 12 | 167.52 |
| 170 | 80 | 3 | 12 | 164.48 |
| 170 | 90 | 2 | 8 | 168.46 |
| 170 | 100 | 2 | 8 | 168.46 |
| 175 | 20 | 7 | 7 | 150.85 |
| 175 | 30 | 6 | 6 | 144.29 |
| 175 | 40 | 4 | 4 | 149.12 |
| 175 | 50 | 3 | 3 | 169.61 |
| 175 | 60 | 2 | 2 | 172.28 |
| 175 | 70 | 2 | 9 | 171.47 |
| 175 | 80 | 2 | 11 | 174.02 |
| 175 | 90 | 2 | 8 | 168.81 |
| 175 | 100 | 2 | 8 | 168.81 |
| 180 | 20 | 7 | 7 | 150.85 |
| 180 | 30 | 6 | 6 | 144.29 |
| 180 | 40 | 4 | 4 | 149.12 |
| 180 | 50 | 3 | 3 | 169.61 |
| 180 | 60 | 2 | 2 | 172.28 |
| 180 | 70 | 2 | 3 | 176.63 |
| 180 | 80 | 2 | 7 | 176.35 |
| 180 | 90 | 2 | 8 | 168.81 |
| 180 | 100 | 2 | 8 | 168.81 |
| 185 | 20 | 7 | 7 | 150.85 |
| 185 | 30 | 6 | 6 | 144.29 |
| 185 | 40 | 4 | 4 | 149.12 |
| 185 | 50 | 3 | 3 | 169.61 |
| 185 | 60 | 2 | 2 | 172.28 |
| 185 | 70 | 2 | 3 | 176.63 |
| 185 | 80 | 2 | 7 | 176.35 |
| 185 | 90 | 2 | 8 | 168.81 |
| 185 | 100 | 2 | 8 | 168.81 |

**Table 1. Test schedule length, simulation effort and max. temperature vs. TL and STCL**